\documentclass[a4paper]{article}
\usepackage{amsmath}
\usepackage{amsfonts}
\usepackage{amssymb}
\usepackage{graphicx}
\usepackage{verbatim}
\usepackage{amsmath, amssymb, amscd, amsthm, amsfonts}
\usepackage{graphicx}
\usepackage{hyperref}
\usepackage{float}
\usepackage{parskip}
\usepackage[left=2cm,right=2cm,top=2cm,bottom=2cm]{geometry} 
\usepackage{caption}
\usepackage{listings}
\usepackage[intoc]{nomencl} 
\usepackage{amsmath}
\usepackage{subcaption}
\usepackage{breqn}
\usepackage{braket}
\usepackage{graphicx, wrapfig}
\usepackage{xcolor}
\usepackage{array}
\usepackage{booktabs}
\usepackage[section]{placeins}
\usepackage{array}
\usepackage{varioref}
\usepackage{svg}
\usepackage{comment}
\usepackage{algorithm}
\usepackage{algcompatible}
\usepackage{algpseudocode}
\usepackage{multirow}
\usepackage{fancyhdr}
\usepackage{mdframed}
\algrenewcommand\algorithmicrequire{\textbf{Input:}}
\algrenewcommand\algorithmicensure{\textbf{Output:}}

\immediate\write18{texcount -tex -sum  \jobname.tex > \jobname.wordcount.tex}

\providecommand{\keywords}[1]
{
  \small	
  \textbf{\textit{Keywords---}} #1
}
\title{Real time QKD Post Processing based on Reconfigurable \\Hardware Acceleration}
\author{Foram P Shingala\textsuperscript{1},
\textsuperscript{*}Natarajan Venkatachalam\textsuperscript{1},
Selvagangai C\textsuperscript{1},
Hema Priya S\textsuperscript{1},
\\Dillibabu S\textsuperscript{1},
Pooja Chandravanshi\textsuperscript{2},
Ravindra P. Singh \textsuperscript{2}}
 
\begin{document}
\maketitle
\begin{center}
    \textsuperscript{\textbf{1}} Society For Electronic Transactions and Security, India\\
\textsuperscript{\textbf{2}}Physical Research Laboratory, Ahmedabad
\end{center}

foram@setsindia.net, *natarajan@setsindia.net, selvagangi@setsindia.net, hema@setsindia.net,\\dillibabu@setsindia.net, pooja@prl.res.in, rpsingh@prl.res.in

\section * {Abstract}
%

Key Distillation is an essential component of every Quantum Key Distribution system because it compensates the inherent transmission errors of quantum channel. However, throughput and interoperability aspects of post-processing engine design often neglected, and exiting solutions are not providing any guarantee. In this paper, we propose  multiple protocol support high throughput key distillation framework implemented in a Field Programmable Gate Array (FPGA) using High-Level Synthesis (HLS). The proposed design uses a Hadoop framework with a map-reduce programming model to efficiently process large chunks of raw data  across the limited computing resources of an FPGA. We
present a novel hardware-efficient integrated post-processing architecture  that offer dynamic error correction, a side-channel resistant authentication scheme, and an inbuilt high-speed encryption application, which uses the key for secure communication. We develop a semi automated High level synthesis framework capable of handling different QKD protocols with promising speedup.  Overall, the experimental results shows that there is a significant improvement in performance and compatible with any discrete variable QKD systems.   \\ 
\newline
\keywords{Hadoop, Key Distillation Engine, HLS, FPGA, classical post-processing, QKD}
\section{Introduction }


Secure data communication is a vital challenge in today’s high-speed networks.  The basic and most critical element of a cryptographic solution is the encryption key, as defined by August Kerckhoffs in 1983\cite{kerckhoffs}. Classical methods of cryptography exploiting the hardness of mathematical problems may sooner be vanquished by the advent of quantum computers. Nevertheless, the same quantum principles that could empower an adversary with enormous computing power can be used to achieve unconditional security in establishing a secret key between two communicating parties, with Quantum key distribution (QKD). There are various implementations of QKD protocols that differ in the way the information is encoded/decoded in quantum states \cite{bennett1992experimental,inoue2003differential,frohlich2017long,wang2021measurement}. In general, it comprises two channels, a quantum communication channel, for transmitting quantum information, and an authenticated classical channel. A quantum communication channel essentially consists of the quantum state of a photon in a particular degree of freedom such as polarization, time bin, phase, frequency, etc., on which the information can be encoded and decoded. The authenticated classical communication channel is used for secret key reconciliation. It is also required to synchronize the transmitter and receiver, separated by large distances. Due to the inherently noisy nature of quantum channels, measurement device imperfections, and improper encoding/decoding, the raw key bits extracted from the quantum channel contain bit-flip errors. On account of the error and the channel losses, post-processing techniques are required to construct the final secret key at both ends.  A robust implementation of QKD, which gives sufficient throughput, in terms of the key rate, in a real-time environment, is challenging. Catering to the speed of quantum communication, a fast key distillation layer or data processing layer, with efficient control hardware, is crucial. \\

The QKD post-processing can be broken down into submodules namely, a) Synchronization, b) Sifting for erasure, c) Sifting for basis reconciliation, d) Random sampling, e) Parameter estimation, f) Information reconciliation (IR) and verification, g) Privacy amplification (PA), h) Key management. These classical components require massive computing and memory resources, hence, were earlier implemented on server systems. But due to complex infrastructure and security assumption of device isolation (trusted node architecture), a solution that is stand-alone, compact, and provides reconfigurability with massive data parallelization, and processing ability, at low power consumption, is the need of the hour. Hence the focus is on FPGA accelerators .\\

Hardware Description Language (HDL) is a specialized language used to describe the structure and behavior of electronic circuits. FPGAs can be programmed using HDL. Any HDL design is directly correlated to resource consumption in FPGA. As the design gets more complex the need to speed the design-flow process makes FPGA developers look at software-based, productivity tools to automate Register-Transfer Level (RTL) design flow. HLS is one such tool and its adaptation to QKD Key Distillation Engine (KDE) is further described in section III. One of the engineering problems that need to be solved in real-time implementations of QKD is the continuous storage and processing of large amounts of data, as quantum encoding and modulation occur at frequencies of hundreds of GHz. FPGA has limited memory storage and management capabilities. Extended memory units like SRAMs and DRAMs can be used along with the FPGA to overcome this drawback. Efficient management and utilization of these additional memory devices can be performed by incorporating a framework like Hadoop to manage big data. This is further analyzed and discussed in section III. \\

In high-performance QKD networks, processing becomes a major bottleneck due to the massive amounts of data collected in the quantum networks, resulting in overhead for the memory and computational capabilities of the targeted systems.  Quantum information reconciliation problems exhibit excessive dynamic computations and memory accesses. Therefore, the overall processing time is dominated by complex computations and unstructured memory management. Throughput and efficiency are the main performance metrics in large-scale quantum key distribution systems. Recently, there is an increased interest to accelerate post-processing using FPGA.  However, performance optimizations and secure processing have not been explored. A technological gap still exists in the practical implementation of a high-throughput and efficient hardware-software co-design for large-scale quantum key post-processing. 
In this paper, we propose Hadoop framework-based FPGA design for high-volume data processing that optimizes memory utilization and at the same time is secure. We report a comprehensive experimental study to evaluate the performance and efficiency using different QKD protocols.  
The Quantum information reconciliation algorithm is implemented using the MapReduce paradigm, in which the error correction process is carried out in a parallel fashion in the FPGA. With any given error correction algorithm and respective constraints, as inputs, the proposed system determines the performance parameters through simulation and then generates the optimized design of the FPGA accelerator. Therefore, any error correction algorithm can easily be implemented using our proposed framework.  We summarize the main contributions of this paper below: 
\begin{itemize}
    \item A hardware-based Key distillation engine with the capability to support multi-protocol discrete variable QKD systems. 
    \item Hardware-based hadoop accelerator to achieve a speedup of the computationally intensive task of information reconciliation and privacy amplification. 
    \item Re-configurable architecture attributing to the framework developed using high-level synthesis technology.
    \item Side-channel attack resistant device authentication algorithm implantation.
    \item Rate – adaptive error reconciliation codes to optimize classical channel throughput, with higher error correction capacity. 
    \item On device, high-speed encryptor with throughput up to 10$Gbps$.
    \item Detailed experimental field trial for three different protocols, namely coherent one-way, BB84, and BBM92.
\end{itemize}

The rest of the paper is organized as follows: Section II covers related work and literature review; section III covers the proposed system design, architecture, and the implementation of KDE in hardware. Section IV defines the experimental setup of the QKD protocols and section V gives the implementation results and the performance analysis of the design and section VI describes the conclusion and future work.
\section{Related Work }
%
One of the first few attempts, in the year 2012, at designing a complete compact QKD system by integrating optics, control hardware, and data processing system into a single chassis, was attempted by Zhang et al \cite{zhang2012real}. The protocol implemented was the decoy-state BB84 protocol. The key distillation software was designed to handle quantum operations at much lower rates as a result of inefficient devices. The Winnow protocol was chosen as an error correction algorithm. The KDE was implemented as a software stack on a computer as part of the integrated design. There have been further advancements in protocol and technology, since. Publishing around the same time, Tanaka, Akihiro, et al \cite{tanaka2012high}, tried to achieve a high-speed phase encoded BB84 QKD system, covering a distance of 50 km, by transmitting with a repetition frequency of tens of GHz using parallel transmission of photons (parallelly connected LED sources) with wavelength Division multiplexing (WDM). Their work also highlights the requirement of large computing and memory resources to be able to derive 1Mb of secure key, using eight XFPs, in Small Form Factor Pluggable (SFP) format, for communication, and multiple FPGA as computing resources for post-processing. Such a mammoth and complicated system architecture would only suffice as a proof of concept. \\

Further, in order to speed up the data processing of measured qubits, individual modules that are part of the post-processing, ought to perform efficiently. The QKD post-processing can be broken down into multiple submodules, each of which is identified to have a specific cryptographic. Multiple teams around the world have worked on all these individual aspects, but a particular article by Cui, Ke, et al. \cite{cui2012real}, aimed at an efficient implementation of the error reconciliation module by exploiting FPGA parallelism and splitting the module such that pipelined execution can be performed between read, write and compute. The team implemented the Winnow error-correcting codes. Based on further studies, it was ascertained that LDPC codes achieved efficiency closest to Shannon’s limit and hence have been prescribed as part of the information reconciliation stack for QKD. \\ 

Early methodical works were carried out for a large-scale project by six research teams in Switzerland \cite{walenta2014fast}. This was a significant step towards a field deployable QKD system. They concentrated on building a QKD system with integrated control hardware and data processing units. The modules for key distillation are described in detail by Constantin, Jeremy, et al \cite{constantin2017fpga}. We have used this work as a reference. The complete firmware was built on a single Xilinx Virtex 6 FPGA. The engine had a post-processing block size of a smaller order. They used a commercial Quantum Random Number Generation (QRNG) from Idquantique and fed the generated random sequence as a seed to a pseudo-random number generator. The COW QKD protocol for distances of 50 km was implemented. \\ 

The alternative to an FPGA accelerated post-processing engine is a pipelined software implementation. Zhou, Jianyi, et al. \cite{zhou2014pipeline} proposed to implement a multi-threaded pipelined approach exploiting more than one CPU core to achieve the optimized arrangement of the major performance parameters. The performance of the pipelined execution is optimized by allowing all stages in the pipeline to have identical processing times. We understand the necessity of parallelization to optimize throughput. Therefore, in our work, we propose to incorporate a parallel data storage and processing framework implemented over the accelerator. In a more recent work by Yuan, Zhiliang, et al. \cite{yuan201810}, a configured host server system with two FPGA-based accelerators, one for QKD control hardware plus sifting and the other for error reconciliation, serves as the post-processing engine in the QKD system. PA of large block size (100Mb) is implemented on the server system using GPU-based parallel architecture. A collective performance throughput of 10Mbps is achieved. The error correction module implemented gives a max correction capacity of 10\%. By adding computational resources like a server, to the architecture, their implementation lacked the main security assumption of an isolated device. This effort was followed by another attempt at improving the throughput of the processing engine by parallelization of the IR phase with larger block sizes of 250 to 350 Kb by Yang et al. \cite{yang2020high}. This post-processing engine was developed for the continuous variable QKD protocol. \\ 

From a review article published by Li, He et al. \cite{li2021fpga}, it is observed that FPGA is an almost mandatory choice for this application and that FPGA also offers an advantage in terms of power consumption \cite{qasaimeh2019comparing}, which can be a key feature for critical applications such as for Satellite Quantum Communication (CubeSat missions) \cite{oi2017cubesat}. This is the first review paper accounting for the work done so far using FPGA accelerators in QKD. The authors also highlight the design productivity gained by using High-Level Synthesis (HLS) to design and configure the accelerator, with features like arbitrary precision arithmetic and parallelization pragmas, etc. Finally moving from FPGA to (system on chip) SoC architecture \cite{stanco2022versatile}, this recent work presents the hardware and software-integrated architecture that can be used in systems that implement practical QKD and QRNG schemes. This architecture fully exploits the capability of an SoC by assigning the time-related tasks to the FPGA and the management to the CPU.\\ 

All the designs up until now have focused on a classical post-processing engine for a specific QKD protocol, implemented either on programmable hardware, software, or GPU and not for a complete, stand-alone, and isolated solution. The focus in the past has been on the algorithmic aspects of the post-processing protocols. In this work, we try to overcome these limitations and give a detailed design and flow of the implementation of the proposed architecture, in the following sections.
\section{System Architecture and Design Methodology}

We propose an FPGA-based flexible, high throughput, and multiple protocol support, distillation engine for QKD systems. The generalized quantum key distillation framework is designed to provide flexibility to adapt to different kinds of QKD protocols. The Key Distillation Engine does all the post-processing and provides the final key to the encryption application. For effective implementation, the entire post-processing flow can be executed in two different phases. 

The preparatory phase, described further in section \ref{CPP}, includes gathering, aligning, and transforming raw data prior to the reconciliation phase. The software components of the preparatory phase are specific to the QKD protocol. In this work, HLS framework is adopted to perform tasks that require modifications with respect to QKD protocol and implementation style.
 In particular, the methods like clock synchronization, measurement alignment, and data sifting are executed in this phase. These strictly depend on the protocol and implementation specifications. The proposed work utilized the HLS framework to create a unified integration platform for QKD post-processing. This platform is depicted in the complete development process flow of the key distillation system in Figure. \ref{design-flow}

\begin{figure}[!ht]
\centerline{\includegraphics[width=0.98\textwidth]{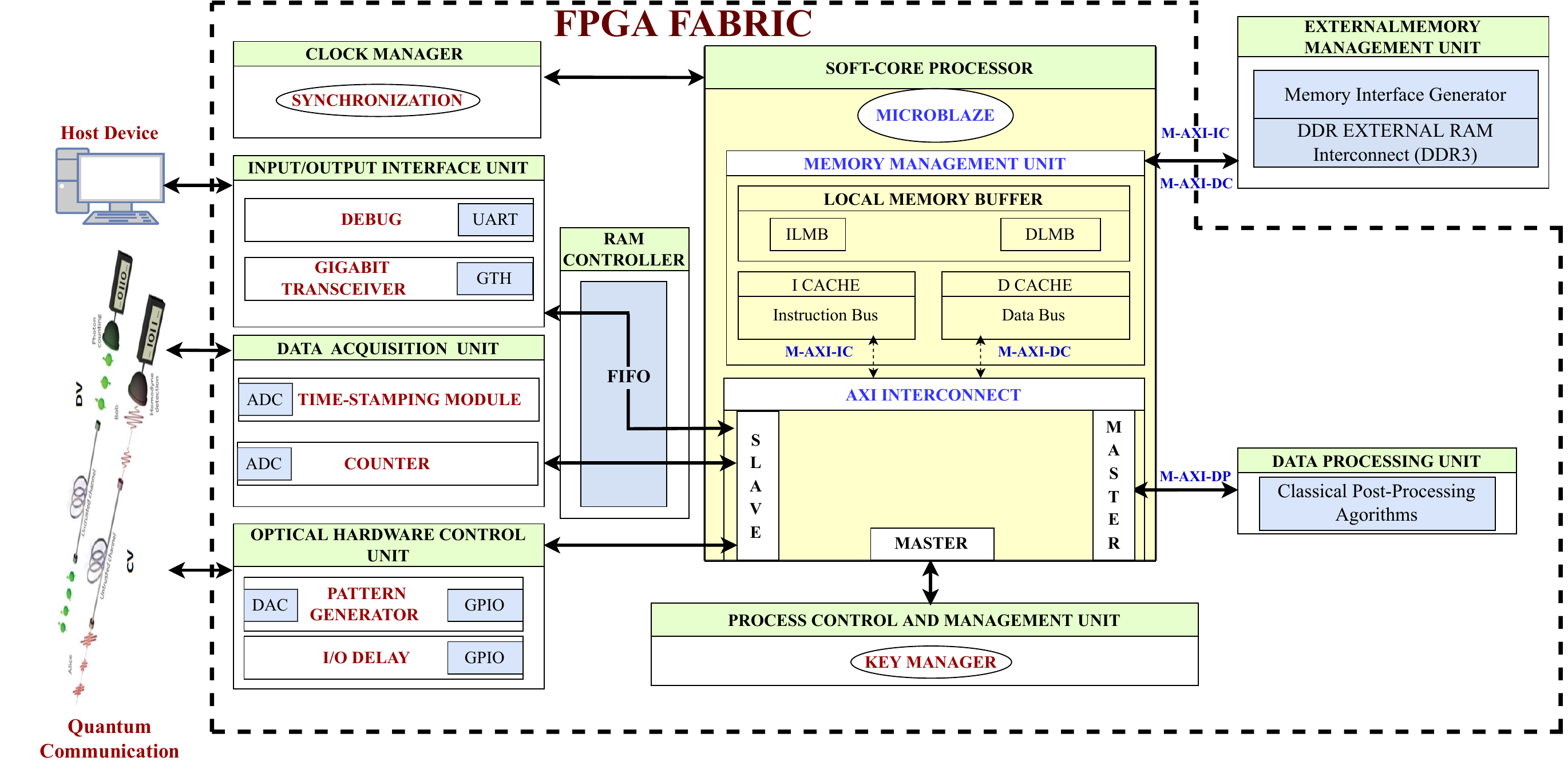}}
\caption{Architecture of the proposed system.}
\label{architecture}
\end{figure}

The reconciliation phase, described in section \ref{P2}, includes the error correction, verification, and privacy amplification modules independent of the QKD protocols. Techniques used in error correction and privacy amplification are computationally intensive. To accommodate this, our hardware design effectively utilizes the Hadoop map-reduce programming model to further optimize the throughput. This helps to attain a trade-off between hardware utilization and fast data processing. The Hadoop data storage and processing framework also aid's in handling the distribution of complex data processing tasks across limited computing resources of the FPGA.   
\textbf{FPGA-Based Processor Design}\label{Archi}\\
The FPGA fabric is designed as a soft-core processor-based System on Chip (Soc) architecture. Process control, task scheduling, and interface for data processing modules are defined as a Software Development toolkit(SDK) API-based software application. Baklouti, Mouna, et al \cite{makni2017hardware} highlights the advantages of an FPGA-based SoC architecture. Figure. \ref{architecture} shows the high-level architecture, which consists of the control unit, data processing unit, and memory management unit.
 The soft-core processor along with the data processing and control modules are implemented on the PL fabric. 1GB DDR3 SODIMM memory is available on board and is used to store data required while processing. The I/O ports used to connect to the host device are USB UART or high-speed serial I/O transceivers (GTH) over Gigabit Ethernet protocol. MicroBlaze is a soft processor core designed for Xilinx FPGAs. MicroBlaze is implemented with AXI interconnect peripheral bus for system-memory mapped transactions with master-slave capability. Each data processing unit module is interfaced with the AXI data port of MicroBlaze for communication. The DDR3 is interfaced with MicroBlaze over AXI instruction cache (IC) and data cache (DC) ports. The framework includes interfaces to and from the optical setup using digital-to-analog converters (DAC) and analog-to-digital converters (ADC), respectively. ADC and DAC are used to control the optical components, e.g. ADC is used to control the synchronous optical laser, and DAC is used for amplitude and phase modulation. The ADC and DAC communicate to the MicroBlaze through data acquisition and optical hardware control units over the AXI- Interconnect. The data acquired from the quantum channel can be buffered in Block Ram (BRAM) units defined as FIFOs. The implementation details of the individual data processing modules and synchronization module are further described in sections \ref{CPP},\ref{P2}, and \ref{poly1305}. The control and data processing units are designed as custom-developed hardware IP blocks, each with a specific function. The software application defines how the control scripts run each custom hardware accelerator.

\subsection{Multi-protocol  QKD Support Workflow}\label{HLS}

Our design approach is directed towards identifying an efficient, configurable, control and process framework that can be incorporated as part of the design solution for any QKD protocol implementation, irrespective of the technology used. This generic framework, for programmable hardware, is designed using high level synthesis(HLS) technology. This technology provides reconfigurability and easy transformation of software description of the algorithm from high-level code into RTL model. The major benefit of HLS is that it provides a platform for individuals that are not experienced in HDL development, to program FPGA \cite{gurel2016comparative}. Vivado HLS is the most popular HLS tool in FPGA Design.\\

\begin{figure}[H]
\centerline{\includegraphics[scale=0.8]{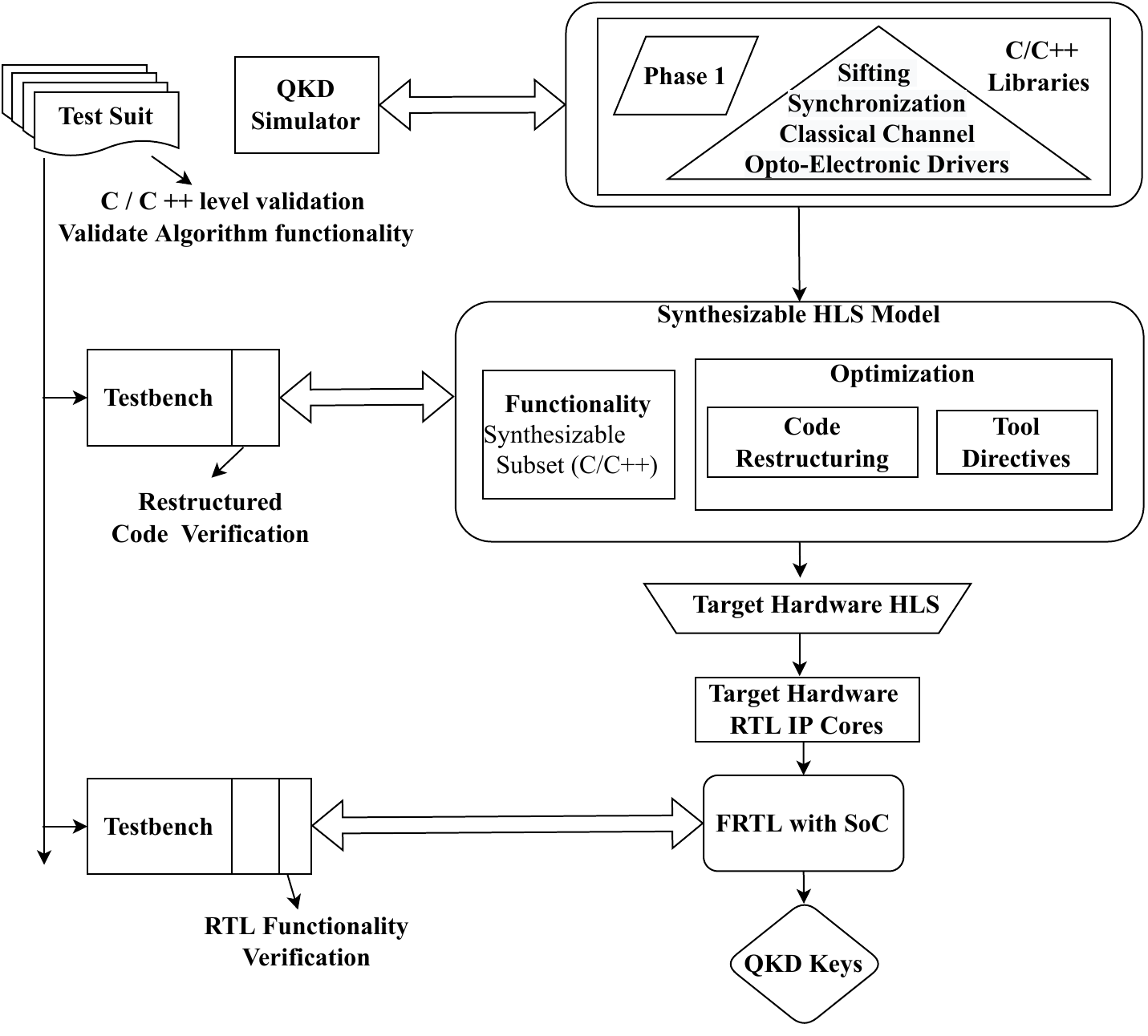}}
\caption{HLS Based design flow for a reconfigurable framework.\cite{umesh2020HLSdesign}}
\label{design-flow}
\end{figure}

Our framework uses the concept of modularity, by including independent IPCores, for each control and data processing task. Control modules can be created by adding software drivers of optoelectronic components (used to perform QKD experiments). These are generally available as open-source libraries in high-level languages (C/ C$++$), and hence can be directly incorporated into our design through HLS. The optoelectronic components are configured to define the parameters (variable attenuation, interference visibility, modulator bias, State of polarization..etc) that help establish a secure quantum channel. Libraries for math and computational utility functions can also be added as part of our design. Figure. \ref{design-flow} shows the design flow adopted by us to develop our data processing modules. The phase I modules or preparatory modules described in section \ref{CPP} are designed after being tested with a QKD simulator. These have to be refined to comply with the synthesizable subset, a test bench has to be written to ensure that the functionality is still intact. A test bench created to validate the algorithm can be used at both the C++ and RTL levels. Optimization directives and datatypes can be added to improve performance. The refined implementation is synthesized using the HLS tool to generate Functional RTL (FRTL).  C functions synthesize into RTL blocks, and function arguments synthesize to RTL I/O while arrays synthesize to memory: RAM or ROM, or FIFO.\\
The benefit of developing in a high-level language is that the control path is implicitly represented, whereas RTL requires the user to explicitly define the data and control paths. For hardware resource optimization, HLS provides user directives. To meet timing constraints, pipelining directives can be incorporated \cite{zwagerman2015high}. Arbitrary precision data types are also provided by Xilinx for vivado HLS \cite{alhamali2015fpga}. Vivado HLS can easily export RTL IP by using “export RTL” option available as part of the tool.\\

\subsection{Preparatory Phase}\label{CPP}

Operation of the control unit follows the QKD protocol configuration and consists mostly of software drivers of the optoelectronic components. Steps to integrate this functionality have been described in section \ref{HLS}. The remaining components of the present key distillation engine can be easily re-configured to any kind of QKD protocol without much effort. All the sub-modules present in the system are designed as a separate proprietary library (IP core) and it can be easily reused for different kinds of QKD setups. Thus the switching time between one protocol to another protocol is decreased, hence making it  easy to build interoperable systems. 

\subsubsection{Synchronization and Classical Channel Communication}

The timing information of the launch of a quantum state from the transmitter’s end and its arrival at the receiver’s end is very crucial in key-sifting and in the establishment of the secret key. We have used the commercial White Rabbit Lite Embedded node (WR-LEN) to achieve time synchronization between the two FPGA boards. WR-LEN is a versatile synchronization solution with any type of classical communication protocol implemented \cite{moreira2009white}. It uses the Synchronous Ethernet (SyncE) and enhanced Precision Time Protocol (PTP) for frequency matching and offset adjustments of the clock. A single channel is a time multiplexed and used for both synchronization and exchange of information for classical post-processing.

The Aurora 8B/10B protocol is a link layer communications protocol for use on point-to-point serial links. Developed by Xilinx, it is used as the classical high-speed(gigabits/second) communication protocol.
The protocol is an open standard and is available for implementation by anyone. Our proposed system supports integration with any standard communication protocol.

\subsubsection{Sifting}
Post-quantum transmission and measurement, Alice and Bob use the classical channel to communicate and derive a secret key. The Alignment step comprises of all the procedures necessary to align the detection times (timestamps) in Bob's reference frame, to Alice's key bits in her reference frame. The effort and the amount of transmitted information can be quite extensive. The protocol used to align is the sliding window protocol and the measurement that confirms the alignment is the correlation coefficient.
\\
The basis sifting procedure filters out any inconclusive or incompatible measurement results at Bob compared to that of Alice's preparation. Depending on the implemented QKD protocol the required information transfer (say measurement basis) might need to be bi-directional as in standard BB84, or uni-directional, as in the distributed-phase-reference protocol. As the output from the sifting procedure, Alice and Bob each hold a set with $n_{sift}$ elements.\\
In the BB84 protocol, Alice stores two bits to describe the prepared state containing key bit value and basis choice. Bob also stores two-bit information on the measurement choice and measurement outcome along with the time-stamp of the detection. For each detection, after the alignment step, Bob announces one bit describing the measurement choice, and Alice responds with one bit containing the XOR value between Bob's and her measurement choice. Sifting can be achieved in multiple ways, but this method is chosen to reduce the communication overhead in the classical channel. Hence, the communication rate for base sifting is $m^{BB84}_{BS}=2*n_{Q}$ bits, where $n_{Q}$ is the number of measurement outcomes Bob has recorded. \\
In the COW protocol, Alice encodes the key information in two consecutive time bins. Alice's $n_{sift}$ elements contain two bits corresponding to each element, to describe the prepared state and an indicator if the state is signal or decoy. Bob's $n_{sift}$ elements contain two-bit information, one for the key bit decoded and the other bit corresponds to which detector clicked i.e. data-line or monitoring detector\cite{stucki2005fast}. Another outcome of the measurement Bob stores is the time-stamp of the detection. For each detection, Bob only announces the $\lfloor timestamp/2 \rfloor$ information (in the alignment phase) and one bit describing the detector click (in the basis sifting phase), while no response is required from Alice to sift out incompatible detections. Hence, the communication rate for base sifting in COW protocol is $m^{COW}_{BS} = n_{Q}$ bits, where $n_{Q}$ is the number of measurement outcomes Bob has recorded.\\
\subsection{Reconciliation Phase}\label{P2}

Information reconciliation is an essential phase in QKD system to arrive at an identical key between both the sender and receiver. Correcting errors in the key material is an important task  and useful for a wide range QKD protocols, as below an error threshold, the errored key can also be used to derive a secure secret key. In this section, we present a novel method for the effective implementation of a computationally expensive part of QKD post-processing stack. We combine error correction, verification, and privacy amplification to adapt the Hadoop map-reduce framework and show the computational benefits. Furthermore, we discuss the hardware realization possibilities which could help to speed up and optimize the efficiency. Additionally, we present  case studies for applying the proposed framework for different types of QKD protocols in order to study the practical
relevance of the proposed techniques.   

\subsubsection{Hadoop Based Accelerator for Reconciliation}\label{LDPC_sec}
\begin{figure}[H]
\centerline{\includegraphics[width=\linewidth]{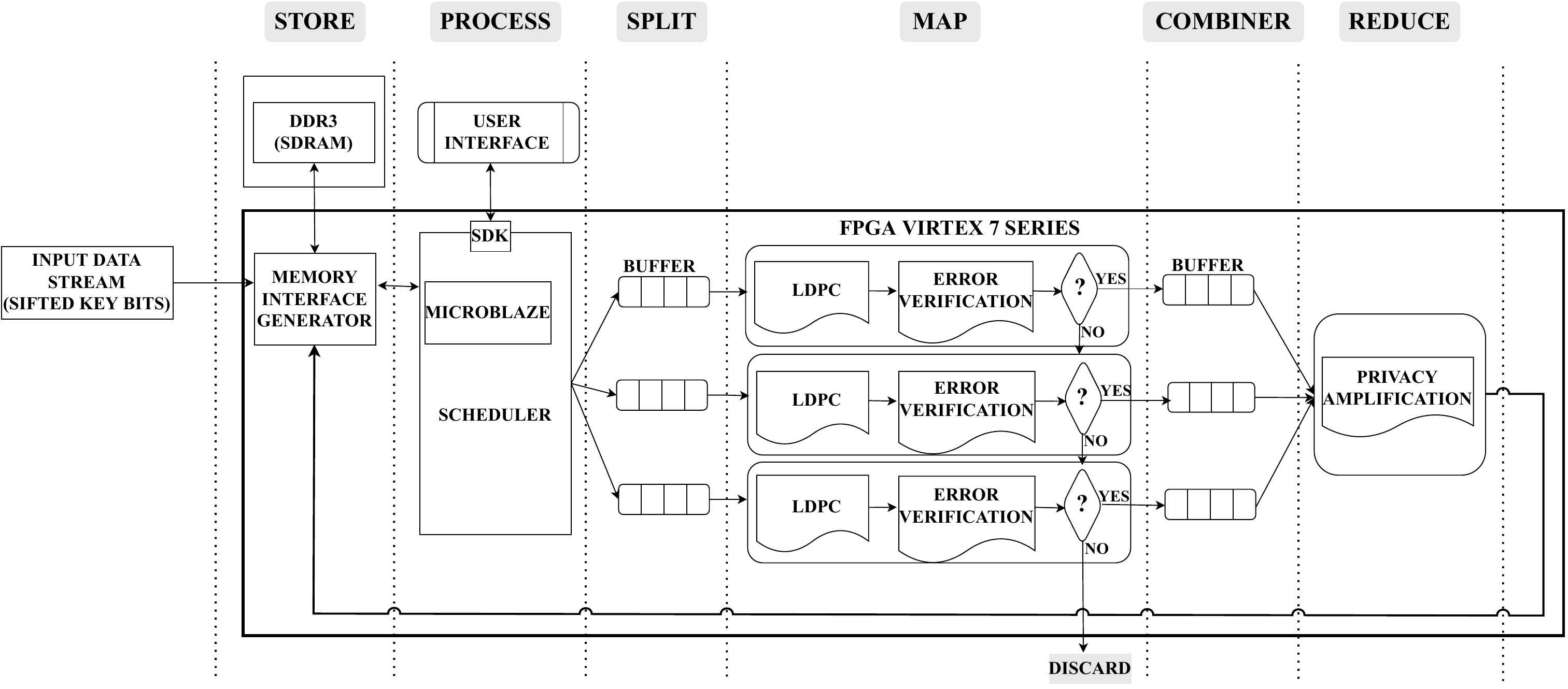}}
\caption{Hadoop-based framework for Information Reconciliation.}
\label{data-flow}
\end{figure}

The proposed design framework incorporates the Hadoop Map-Reduce programming model to efficiently store and process large chunks of sifted key bits obtained from the quantum setup, at a rate faster than the FPGA’s clock. To overcome a possible bottleneck at this point we use a mechanism to buffer the incoming data and parallelly process multiple blocks of sifted key bits simultaneously. Map Reduce programming model comprises of Splitting, Mapping, Combining, and Reducing processes.
The flow of data across these stages is depicted in Figure. \ref{data-flow} and the Algorithm \ref{algimpl} captures the implementation aspect of the framework.\\
Input data is stored in an onboard memory SDRAM (DDR3) and then accessed by the scheduler as blocks of fixed size. The scheduler runs a $C++$ SDK-based application that controls and interconnects all the data processing and control modules of the KDE. The input data is split into blocks by the scheduler with respect to user-defined block size and fed to the Mapper. Mapper here is implemented as the IR and verification module. Due to finite key security effects, the input block size for PA is chosen to be large $(10^6 bits)$. Being limited by the resource-intensive IR step for larger block sizes, multiple iterations of IR with smaller block sizes is the chosen design approach.  Multiple instances of the mapper run simultaneously. The output of the mappers is temporarily stored in the Block Ram and is then combined and sent to the PA module, which here acts as the reducer. Incorporating a map-reduce framework in KDE has helped achieve high throughput by parallelizing the KDE modules with respect to the block size. 
Raw key obtained by quantum transmission is first sifted. The error introduced while transmission of single photons (channel transmissivity) is then estimated. This information drives the selection of the IR code parameters, which is explained in section \ref{ldpc}.  Post this the scheduler (MicroBlaze) invokes the map-reduce framework. After the PA step, the final key is written back to the DDR3 memory unit. 
The inspiration to incorporate the Hadoop framework into our design model was derived from the implementation by Neshatpour, Katayoun, et al \cite{neshatpour2018energy}.
    
\begin{algorithm}[H]
 \caption{Combined Error correction and Privacy Amplification process using map-reduce framework}\label{algimpl}
\begin{algorithmic}
\Require Siftedbits, Bit Error Rate (BER), security parameter
 \Ensure  Final key 
 \\
  \textit{Initialisation} :  Matrix $[H]_{m x n}$, $BER =0$, security parameter $\rho$
  \State \textit{class} \textbf{Mapper}  
\\  \textbf{Method} Map (Sifted\_bits, BER, LDPC Hadoop Instance)
  \For {All blocks} 
   \State  Error_correction (Sifted\_bits, Blocksize)
  \If {$BER = 0$}
  \State  Emit (Error\_Corrected\_Key) 
  \EndIf
  \EndFor
  \For{All blocks} 
  \State Error_verification
  \If {(Result=success)}
  \State Emit (Error\_Corrected\_Key, Result)
  \EndIf
  \EndFor
    \State \textit{class} \textbf{Combiner} 
   \\ \hspace{0.2cm} \textbf{Method} Combine (Error\_Corrected\_Key, LDPC Hadoop Instance)
   \State \hspace{0.2cm} Emit (Result)\\      
    \hspace{0.2cm} \Return  Combined\_Corrected\_Key
    \State  \textit{class} \textbf{Reducer} 
   \\ \hspace{0.2cm} \textbf{Method} Reduce (Combined\_Corrected\_Key, security parameter)
  \State \hspace{0.2cm} Privacyamplification()
    \State \hspace{0.2cm} Emit (Result)  \\    
    \hspace{0.2cm} \Return FinalKey
\end{algorithmic}
\end{algorithm}

\subsubsection{{\textbf{Parameter Estimation}}}

Once Alice and Bob have exchanged timestamps and sifted the key bits, the transmitter tries to estimate the approximate lower limit of the error introduced while transmission of the quantum states. This is done by checking with reference to a random subset of the sifted key bits (extracted by random sampling) shared by the receiver. 
\begin{center}
\textit{Error rate of Sampled subset $\approx$ Error rate of remaining bits + $\Delta$ }
\end{center}
The above inequality is proved using Chernoff - Hoeffding type bounds and depends on the sample size. Random Sampling is done using an FPGA-based True Random Number Generator (TRNG)). After randomly sampling a subset, these bits along with their time stamps are exposed over the classical channel and hence are discarded by both parties. The approximate error rate is estimated by,
\( \sum_{n=1}^{r} \frac{1}{N} \), where $r$ is the number of errored bits received  and $N$ is the total number of exposed bits.
This value is captured in the QBER of the quantum channel. In the proposed design, the random sampling of exposed bits and the QBER calculation is implemented on the MicroBlaze processing system of the FPGA as it requires modulo 2 additions and a single 32-bit division operation. The QBER estimate plays a major role in determining if a secure secret key can be extracted by further processing. If the QBER is above a given threshold (defined for each protocol, depending on the errors due to device and measurement imperfections, and on the amount of information that can be leaked to an all-powerful quantum adversary through the devices or the channel) the iteration is aborted and the key derived is discarded. \\

\subsubsection{\textbf{LDPC Error Correction}}\label{ldpc}

\begin{figure}[H]
    \centering
    \includegraphics[scale=0.7]{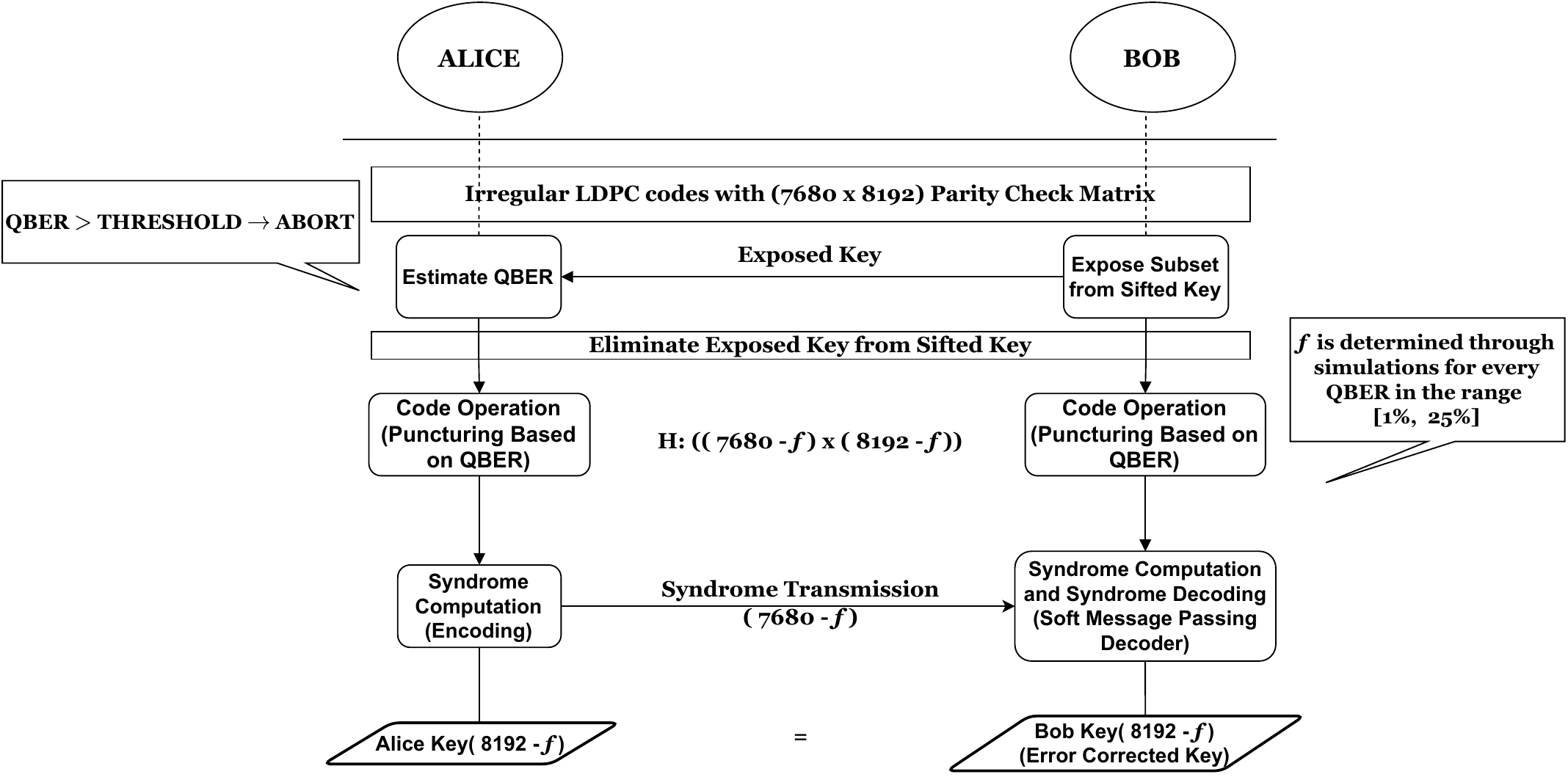}
    \caption{The flow diagram of the LDPC codes}
    \label{LDPC}
\end{figure}

Recently, LDPC codes have been researched extensively, from among the family of Forward Error Correction Codes for QKD \cite{martinez2014demystifying,elkouss2010information,dixon2014high}. LDPC codes are implemented as an IP core using the HLS design flow described in section \ref{design-flow}.  The technique used to construct the parity check matrix is protograph code construction. Protograph codes are constructed by expansion of a base protograph. The resulting LDPC parity check matrix is a combination of sub-matrices. The proposed system implements an irregular parity check matrix populated with elements from the GF(2) field and a soft-decision
message-passing decoder. Index positions of the elements of the matrix containing a one are stored in the local memory of MicroBlaze. The row and column indexes are used to construct a tanner graph at the decoder to iteratively decode the syndrome and a strong belief is derived for the received key bits. The encoding and decoding technique used in the proposed design is a standard technique described in literature \cite{elkouss2010information,dixon2014high}. 
The QBER also helps us in selecting an efficient LDPC code from the ensemble of codes, with the dimension of the given code defined such that the limit derived through Shannon’s channel capacity theorem is achieved to the maximum possible extent. This is established by determining the threshold of the LDPC code using the measured QBER. Andrew Thangaraj et al. \cite{pradhan2015construction} elaborates more on this and we refer to this work as we have derived the LDPC codes for the proposed design from their work. \\
Figure. \ref{LDPC} describes the algorithmic flow of the implementation. 
Based on the QBER one can quantify the amount of extra information which is to be added or extracted to reconcile the errors. Refer to the work by Elkouss, David et al. \cite{elkouss2010information} for more details. These are termed as rate-adaptive LDPC codes and are required to withstand the variation in QBER. Based on Monte Carlo simulations, the codes chosen for the proposed work have the capacity to correct 25\% block errors with a maximum tested input size of $10^9$ bits (1 Gb) and the maximum decoder iterations set to 50. The technique used to optimize code rate and achieve maximum channel capacity is known as shortening of message bits and puncturing of parity bits, which is covered in detail in \cite{elkouss2010information}. In our proposed work we have used the above techniques and reduced the dimension of the parity check matrix to $((7680-f) x (8192-f))$, $f$ here is the reduction factor. Take a full-capacity LDPC code and reduce it based on the QBER. The Monte Carlo simulations were run for error rates from the range of 15\% to 25\% to identify this reduction factor. 

\subsubsection{\textbf{Error Verification}} 

 Error verification is an essential step after error correction, for the integrity check of the decoded key bits. There is a finite chance that the sifted key bits shared by Alice and Bob still have some error, and an additional step of error verification is necessary. To implement this integrity check, the same Universal Hash Function \cite{carter1979universal} technique, which is used for authentication, described in  section \ref{poly1305},  can be reused. The error-corrected key will be the input to the hash function along with a pre-shared key, which can be generated from the TRNG module or as part of the QKD-generated key from the previous iteration. Based on our implementation, the block size of the input is approximately $10^6$ bits. To further ensure information-theoretic security, a pre-processing procedure is followed as shown in Algorithm \ref{algcap}, to extract 128-bits, required as per the architecture of Poly 1305 algorithm, from a large chunk of input. The error corrected key (N-Bit) is divided into 16 chunks labeled as (C1, C2….C16) each of size N/16 bits. Further, each chunk is equally divided into two sub-blocks labeled as {(D1, D2) … (D31, D32)}, of size N/32 bits, and the XOR of the two sub-blocks is evaluated and the result of which is again divided into two sub-blocks and the procedure is repeated until we arrive at an output of 128-bit. This output associated with each chunk ((C1, C2….C16)) along with the 256-bit pre-shared key is given as input to poly1305, which will generate a 128-bit tag. The tags generated by both Alice and Bob are compared at Bob. Bob will then share the error verification flag with Alice. If any of the tags are erroneous, the specific chunk will be discarded. 

\begin{algorithm}[H]
\caption{Error Verification Algorithm }\label{algcap}
\begin{algorithmic}
\Require Error corrected key of size N.
\Ensure \hspace{0.02cm} 16 Tags each $128$ bits.
\vspace{0.2cm}
\State $t\ \gets\ 16$
\State $k\ \gets\ 128$
\State $Eck\ \gets\ Error\_Corrected\_Key[b_1,\cdots,b_N]$
\vspace{0.05cm}
\State $[C_1,C_2,\cdots, C_t] \gets Split(Eck)$
\vspace{0.05cm}
\State $Size \gets \frac{N}{t}$
\vspace{0.05cm}
\While{$Size > k$ bits}
\vspace{0.05cm}
\For {\texttt{$i \gets 1,t $}}\Comment{Iterate over each chunk}
\vspace{0.05cm}
    \State $D_{2i-1},D_{2i}  \gets Split(C_i)$\Comment{$ D_i, size =size/2$}
    \vspace{0.05cm}
    \vspace{0.05cm}
    \State $C_i \gets XOR(D_{2i-1}   ,D_{2i})$  
    \vspace{0.05cm}
\EndFor
\State $Size \gets \frac{Size}{2}$
\EndWhile
\vspace{0.05cm}
\State $Tag_i \gets \textbf{Poly1305} ( C_i  ,random $ $seed))$
\end{algorithmic}
\end{algorithm}

\subsubsection{\textbf{Privacy Amplification}} 

Privacy amplification (PA) is one of the foremost vital post-processing procedures of Quantum Key Distribution (QKD). Within the post-processing part, the transmission of key bits from Alice to Bob requires mandatory communication on the service channel for sifting, error correction, and detection, leading to information leakage, which can also occur due to eavesdropping attacks. Hence, to ensure an exact level of security, it's necessary to get rid of this amount of leaked information through a privacy amplification step. In privacy amplification, the partly secure string is to be reworked into an extremely secret key by public discussion. It's been shown that this could be done by computing the output of a random, however publicly chosen two-universal hash function, applied to the input string, resulting in a secret and secure key. For this, we use Toeplitz hashing to cut the error-corrected bit stream by the adjustable compression ratio, guaranteeing the safety of the remaining secret key bits.
The simplest implementation idea of a large-scaled PA scheme is directly performing multiplication operations between secure key W and Toeplitz matrix T(A), resulting in the computational complexity of \textit{O($n^2$)}. We reduce the complexity to \textit{$O(nlogn)$} by performing fast Fourier transform (FFT) instead of normal multiplication \cite{constantin2017fpga}.

Weak secure key W with a length of $n$ from the basis key sifting and error correction procedures. Then, Alice and Bob decide on the final secure key length r with rigorous statistical analysis. Further, Alice and Bob publicly discuss a random seed with a length of $(n-1)$ bits to construct the universal hash function. The random seed is generated using an FPGA-based TRNG on Alice's side.

Our HLS-based PA scheme for KDE mainly consists of three steps: splitting and shuffling, sub-PA, and secure-key merging.

\subsection{Authentication}\label{poly1305}
We make use of Wegman-Carter framework based authentication scheme for message integrity and user identity verification over the classical channel. Wegman-Carter framework \cite{carter1979universal} based authentication is proven to be information-theoretically secure. For achieving a practical authentication we need a universal hash family which will then have to be combined with a pseudo-random function to get a MAC. Universal Polynomial hashing is said to be highly collision resistant. Bernstein found a special prime number of the form ${2^{130}-5}$ for working with 128-bit coefficients  and proposed a new MAC defined in equation \ref{eq:2}, Poly1305-AES \cite{bernstein2005poly1305}. The original proposal of Carter and Wegman for a universal family was to pick a prime $p\  \geq\  m$, m be an integer, 
and then define 
\begin{center}
\begin{equation} \label{eq:1}
  h_{a,b}(x)\   =\  ((ax + b) mod \ p) mod\  m
  \end{equation}
  
\end{center}

Commercial QKD systems use a universal family based on polynomial evaluation. 
We use an implementation-friendly adaptation of their proposal following the work of Bernstein. The basic idea is to parse the message into 16-byte chunks which form coefficients of a polynomial and evaluate it at r(key) modulo a suitable prime number.

\begin{center}
\begin{equation} \label{eq:2}
 tag(t) := Poly1305 \oplus PRF  := h_{k_1}(m)\oplus k_2 
  \end{equation}
   \end{center}

\textbf{Horner's method} is the most efficient and hence recommended implementation for  polynomial evaluation due to the repeated multiplication with a fixed (secret) multiplicand. A polynomial of the form $ p(x)$ is evaluated at $k$ using Horner's method as given in equation \ref{eq:3} 
\begin{center}
\begin{equation} \label{eq:3}
 p(k)= a_0+ k (a_1 + k (a_2 +\ldots k(a_{n-1}+ a_nk))). 
  \end{equation}
   \end{center}
 Hence Horner's method requires only $n$ multiplications instead of $n(n+1)/2$ multiplications needed by the naive method.
\\
\subsection{AES Encryption}
An efficient hardware architecture design of Advanced Encryption Standard (AES-128) is implemented. The AES algorithm as defined by the National Institute of Standards and Technology (NIST) of the United States, has been widely accepted. The throughput of our implementation is beyond 10Gbps for the encryption and decryption process with device XC7VX485T of the Xilinx Virtex Family. The hardware design approach is entirely based on pre-calculated look-up tables (LUTs) and parallelly executable instances which results in a less complex architecture, thereby providing high throughput and low latency. The speedup achieved is by running the key expansion module independent of the AES rounds. 
AES with a larger key size is considered resistant to attack by a powerful quantum adversary and hence is chosen as the application that uses the QKD-generated key, thus carrying out secure image, video, or message encryption.

 \section{Experimental Setup}
\subsection{Description}
\begin{figure}[H]
\includegraphics[width=0.5\textwidth]{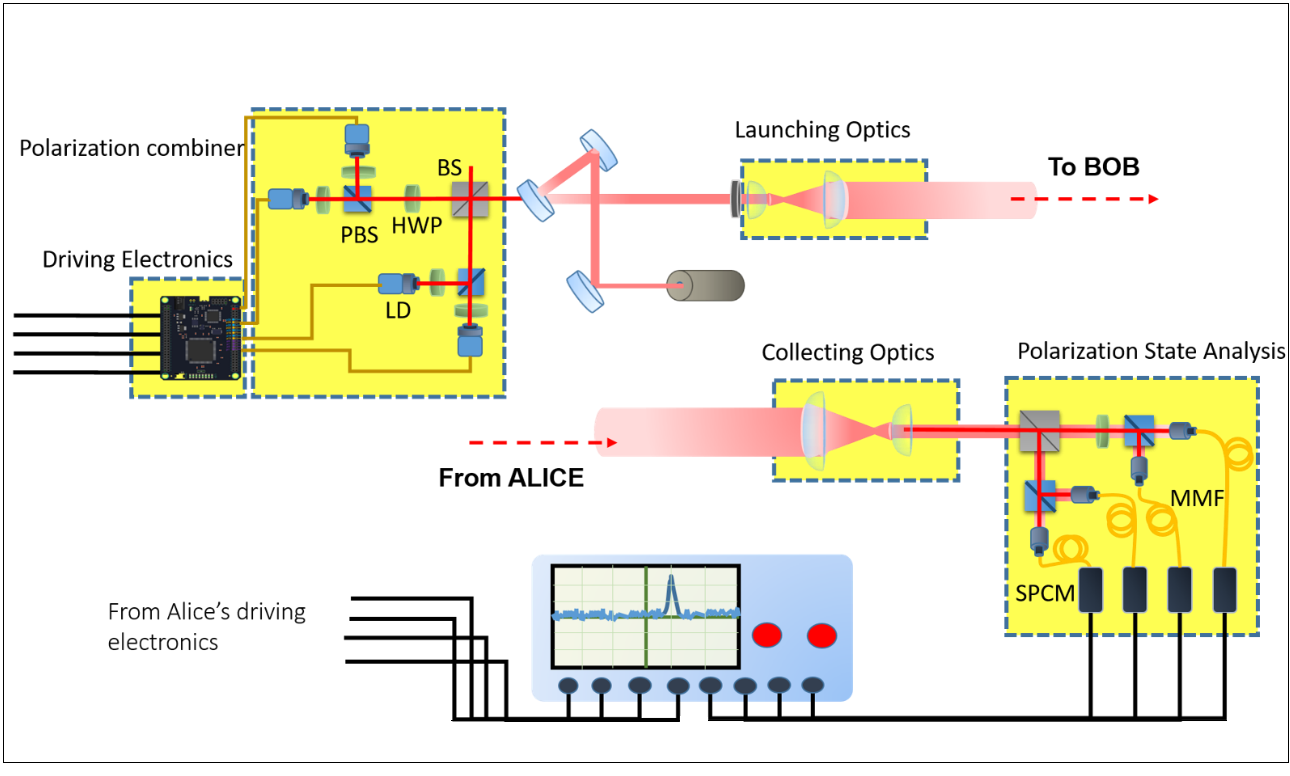}
\includegraphics[width=0.5\textwidth]{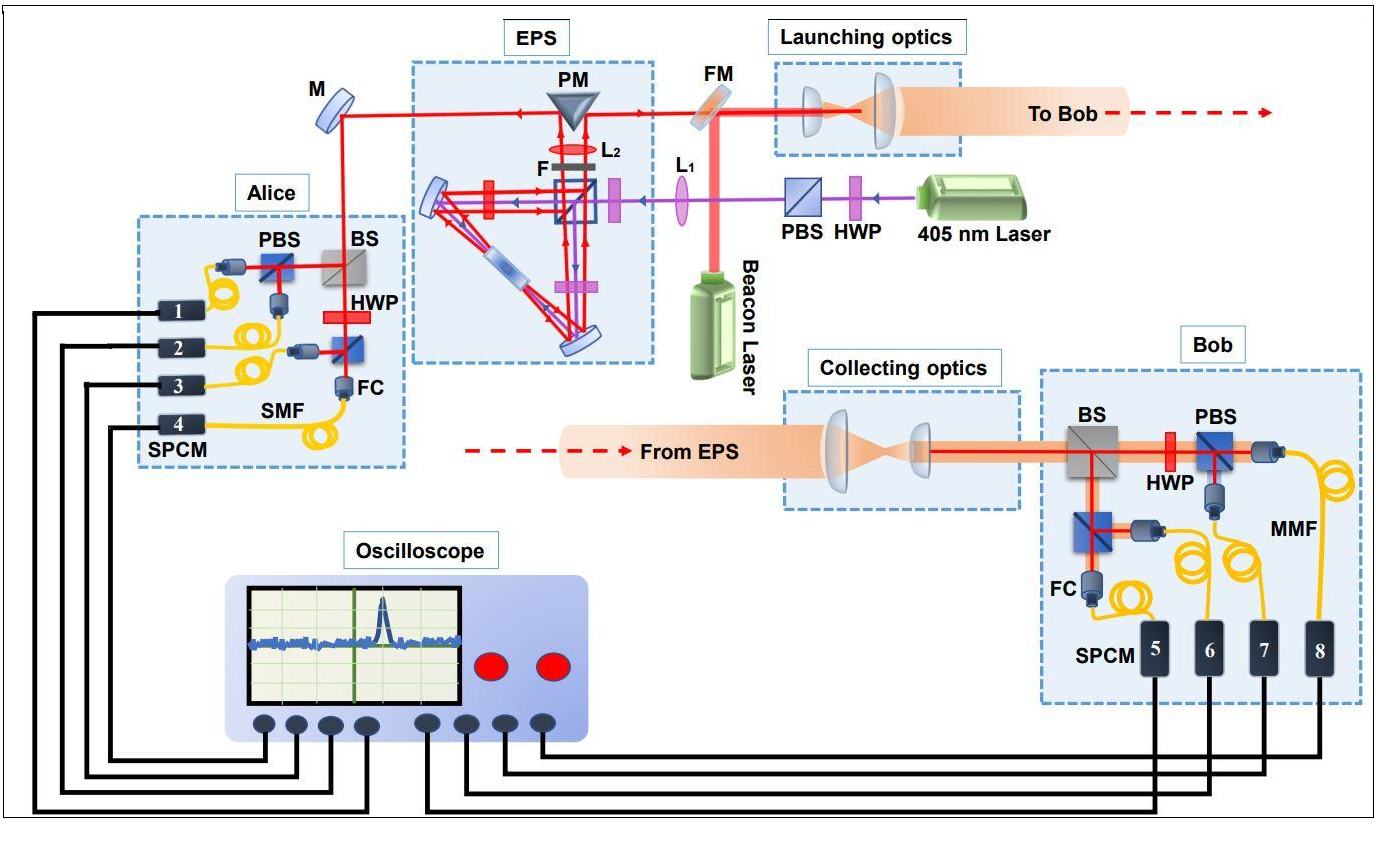}
\caption{Experimental scheme of $(a)$ BB84 protocol over 200 meters and of $(b)$BBM92 protocol, that includes both optical and electronic arrangements.\\EPS: entangled photon source, FM: flip mirror, PM: prism mirror, M: mirror, F: filter, FC: fiber coupler, BS: beam splitter, PBS: polarization beam splitter, DPBS: duel wavelength PBS, HWP: half-wave plate, SMF: single-mode fiber, MMF: multi-mode fiber, SPCM: single-photon counting modules, PPKTP: Periodically poled potassium titanyl phosphate.  LD: laser driver, BD: beam dumper.}
\label{expt}
\end{figure}

To validate and analyze the performance of the key distillation engine designed and implemented, data is collected from quantum experiments performed at Physics Research Laboratory, Ahmedabad, implementing polarization-based BB84 QKD protocol \cite{biswas2022quantum} and BBM92 QKD protocol.  The details of the experimental setups are shown in Firgure.\ref{expt}. In the BB84 setup, Figure. \ref{expt}(a), weak coherent pulses are generated by using a variable optical attenuator at the output of a pulsed laser with a repetition rate of 80 MHz. The encoded state is then propagated in a free space lossy medium with channel transmissivity estimated at 70\%. At Bob’s end, there is a polarization-based detection setup consisting of a balanced beam splitter (passive random basis selector) with a polarizing beam splitter (PBS) on the reflected arm (measurement in ${H,V}$) and a combination of the half-wave plate with PBS (measurement in ${D, A})$ at the transmitted arm. Photons at the output ports of the PBS are detected by fiber-coupled avalanche photodiodes (Excelitas \textit{SPCM AQRH-14-FC}). BBM92 protocol is just the entangled version of the BB84 protocol. BBM92 protocol involves pairs of entangled photons. In this protocol, a common sender (EPS) prepares the entangled photon source (EPS) and sends them to Alice and Bob through the quantum channel. In Figure. \ref{expt}(b), the Polarization Sagnac interferometer is used to prepare entangled photons. In this interferometry, a diagonally polarized $405 nm$ continuous-wave laser with an output power of $\sim 5 mW$ is used to pump a 30 mm long Type-0 PPKTP crystal of period $3.425 \mu m$. A lens L1 of focal length $400 mm$ is used to focus the pump beam on the crystal to generate entangled photons. The horizontally polarized pump beam is transmitted through DPBS in a clockwise direction, and vertically polarized light is reflected through the DPBS in a counter-clockwise direction. Since both the clockwise and counter-clockwise pump beams follow the same path but in opposite directions inside the Sagnac interferometer and the \textit{Type-0} PPKTP crystal is placed symmetric to the DPBS, the implemented scheme is robust against any optical path changes to produce SPDC photons in orthogonal polarizations with ultra-stable phase. At the output of the Sagnac interferometer, a filter (F) is used to block the pump beam while transmitting the entangled photons. A prism mirror (PM) is used to separate the entangled photon pairs. One photon is sent to Alice, and another photon is sent to Bob through launching optics. The detection setup is the same as BB84. The output from the SPD is fed into electronics for recording the counts per integration time and this data is then used to derive the sifted key bits. The sifted key bits are the input to the key distillation engine.   

\section{Implementation and Analysis}
\subsection{Security Analysis of the Implementation}

Instead of trying to break the theoretical foundations of a given cryptographic system (which is proved to be unconditional), another “attack philosophy” is to attack its implementation via loopholes, in order to gain some secret information via unconventional channels. A practical implementation of QKD protocol is never perfect and the performance of the protocol depends on the applicability of the security proofs and assumptions to the real devices \cite{bacco2013experimental,xu2020secure}, as well as on numerous parameters, including post-processing efficiency and the level of noise added to the signal at each stage (including the noise added due to attenuation). These assumptions include the existence of an authenticated channel between Alice and Bob, the isolation of the trusted devices (i.e., that Eve cannot access Alice and Bob’s devices), and that the devices perform in the way that they are expected to. Exploitable imperfections in the trusted party, that allows quantum hacking, are called side channels. One such side channel was identified by performing the side-channel analysis (SCA) and a countermeasure was proposed by us. The authentication module, implemented as part of the QKD key distillation engine, is assessed to be a side-channel resistant implementation on hardware. Side channel analysis of these classical components contributes significantly to the security model of the QKD system. Due to the composability of QKD, the security of the QKD system depends on the security of all its other constituent components.
\subsubsection{Side-Channel Attack on Implementation}
\begin{figure}[H]
        
 \includegraphics[width=0.495\textwidth]{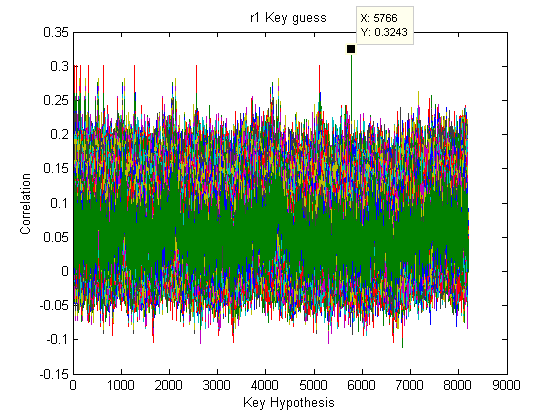}
\includegraphics[width=0.495\textwidth]{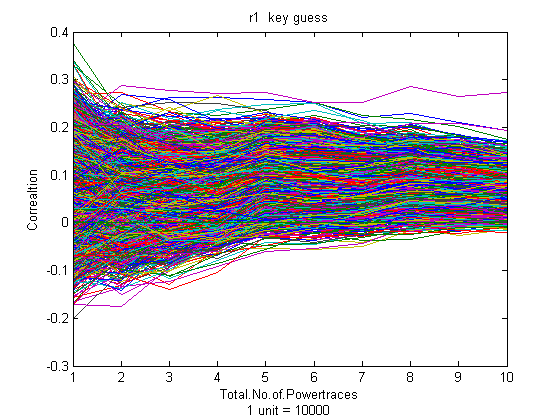}
\caption{Correlation Power analysis of Poly1305}
\label{CPAPoly1305}
\end{figure}

Correlation power analysis of FPGA is used to reveal the first 13-bit key part $k_1$, for that 13-bit key hypothesis is required. Therefore, the attack complexity required to recover the key, $k_1$ is $2^{13}$. We recorded one lakh power traces as shown in the Figure. \ref{CPAPoly1305} to derive the 13-bit key part.
\noindent\textbf{Attack Complexity: } Thus the total attack complexity amounts to 5* $(2^{13})$+ 1*$(2 ^ {12})$ + 1*$(2 ^ 8)$ +    3* $(2 ^ 7) $ $\approx$ $6*(2^{13})$ $\approx$  $2^{15}$.  
Therefore, the overall attack complexity to retrieve the entire key is about $2^{15}$.\\
\textbf{Implication: }  $h_{k_1}(m)\oplus k_2 := t$ where $h$ is a universal hash, $k_1$ is fixed and $k_2$ is one-time key. For encrypting each tag, the one-time-pad key is replaced with a part of the key produced in the previous QKD session. Note that the described attack gives $k_1$ and $(m,t)$ is public. Thus one can compute $k_2=h_{k_1}(m)\oplus t$ and hence, forgery or MITM(Man-in-the-Middle attack) is possible by computing $t'=h_{k_1}(m')\oplus k_2$ for any desired message. In fact, commercial systems from IDQuantique, use polynomial evaluation-based universal hashing for authentication and thus appear to be susceptible to the attack.\\
\noindent\textbf{Countermeasure: }
The proposed MAC construction is as follows. 
Using the universal hash function, a new key, $k_1^{new}$ has to be generated for each authentication as shown in the equation \ref{eq.4}.  $k_1^{new} := T_r * k_2$\\
By selecting a random bit string, r = ($r_1, r_2,\cdot,r_{N+L-1}$) of $N+L-1$ bits, a Toeplitz matrix $T_r$ is constructed, which is multiplied by vector component, one time key $k_2$ to get $k_1^{new}$. The random bit string is generated using TRNG.

 \begin{center}
 \begin{equation}\label{eq.4}
     k_1^{new} = \begin{bmatrix} r_L&r_{L+1}&r_{L+2}&...&r_{N+L-1}\\
r_{L-1}&r_{L}&r_{L+1}&...&r_{N+L-2}\\
...&...&...&...&...\\
r_{2}&r_{3}&r_{4}&...&r_{N+1}\\
r_{1}&r_{2}&r_{3}&...&r_{N}\\
\end{bmatrix} \begin{bmatrix} k_2^1\\
k_2^2\\
...\\
\dots\\
k_2^N\\
\end{bmatrix}
 \end{equation}

\end{center}
Then protected tag can be generated by equation \ref{eq.5}
\begin{center}
    \begin{equation}\label{eq.5}
        t := h_{{k_1}^{new}}(m)\oplus k_2 .
    \end{equation}
\end{center}

\subsubsection{Implementation Result}
The developed Hadoop-based reconfigurable KDE hardware design is tested with quantum bits (raw key bits) obtained from an experimental setup of the BB84 protocol, BBM92 protocol, and COW protocol at the Physics Research Laboratory (PRL), Ahmadabad. The results obtained are recorded in the below tables. Table.\ref{uti} records the area and utilization parameters for the hardware implementation of our design, and Table.\ref{impl} records the performance of our implemented design in various experimental settings. The designs are implemented on a Virtex-7 VX485T Xilinx FPGA.

\begin{table}[hbt!]
\renewcommand{\arraystretch}{1}
\centering
 \begin{tabular}{|p{4.3cm}|p{1.5cm}|p{2.7cm}|p{2cm}|p{2cm}|}
	\hline 
	\textbf{\centering{DESIGN}} & \textbf{\centering{LUT\%}} & \textbf{\centering{FF(Flip-flop\%)}} & \textbf{\centering{BRAM(\%)}} & \textbf{\centering{Total Power(W)}} \\ 
	\hline
	Without Hadoop framework (Transmitter)	&9.14	&\centering{4.50}	&23.6	&3.286 \\ 
		\hline
    With Hadoop framework (Transmitter)  &9.71   &\centering{4.91}   &38.7	&3.525 \\ 
		\hline
    Without Hadoop framework (receiver)  &11.99  &\centering{5.56}   &32.2	&3.837 \\ 
		\hline
    With Hadoop framework (receiver)  &18.46  &\centering{8.26}	&61.2	&5.131  \\ 
		\hline
\end{tabular} 

\caption{Utilization report of the integrated Key Distillation Engine design with and without hadoop framework on VC707 dev board for both transmitter and receiver. \label{uti}}
\end{table}

\begin{table}[hbt!]
\renewcommand{\arraystretch}{1}
\centering
\begin{tabular}{|>{\centering}p{1.8cm}|>{\centering}p{2.2cm}|>{\centering}p{1.6cm}|>{\centering}p{2.4cm}|p{2cm}|p{2cm}|}
\hline
\textbf{Protocol} & \textbf{Input Block size (bits)} & \textbf{QBER} & \textbf{No of parallel Instances} & \textbf{Key Rate (Kbps)} & \textbf{Time \hspace{2cm}(seconds)} \\ 
\hline
{}{} & 10,07,616 & {}{}{25\%} & 1 & 39 & 25.7 \\ 
\cline{2-2}\cline{4-6}
 & 10,07,616 &  & 3 & 116.7 & 8.6 \\ 
\cline{2-2}\cline{4-6}
 & 9,83,040 &  & 4 & 157.9 & 6.2 \\ 
\hline
{\textbf{BB84}} & 10,07,616 & {}{}{2.63\%} & 1 & 96.8 & 10.37 \\ 
\cline{2-2}\cline{4-6}
 & 10,07,616 &  & 3 & 285.1 & 3.52 \\ 
\cline{2-2}\cline{4-6}
 & 9,83,040 &  & 4 & 368.1 & 2.66 \\ 
\hline
{}{}{\textbf{COW}} & 10,07,616 & {}{}{21.40\%} & 1 & 45.4 & 22.13 \\ 
\cline{2-2}\cline{4-6}
 & 10,07,616 &  & 3 & 128.7 & 7.8 \\ 
\cline{2-2}\cline{4-6}
 & 9,83,040 &  & 4 & 175.8 & 5.57 \\ 
\hline
{}{}{\textbf{BBM92}} & 10,07,616 & {}{}{9.03\%} & 1 & 70.6 & 14.22 \\ 
\cline{2-2}\cline{4-6}
 & 10,07,616 &  & 3 & 207.4 & 4.84 \\ 
\cline{2-2}\cline{4-6}
 & 9,83,040 &  & 4 & 273.5 & 3.58 \\
\hline
\end{tabular}
\caption{Implementation results for the Hadoop framework for multiple instances of the mapper.\label{impl}}
\end{table}

\begin{figure}[H]
 \includegraphics[width=0.5\textwidth]{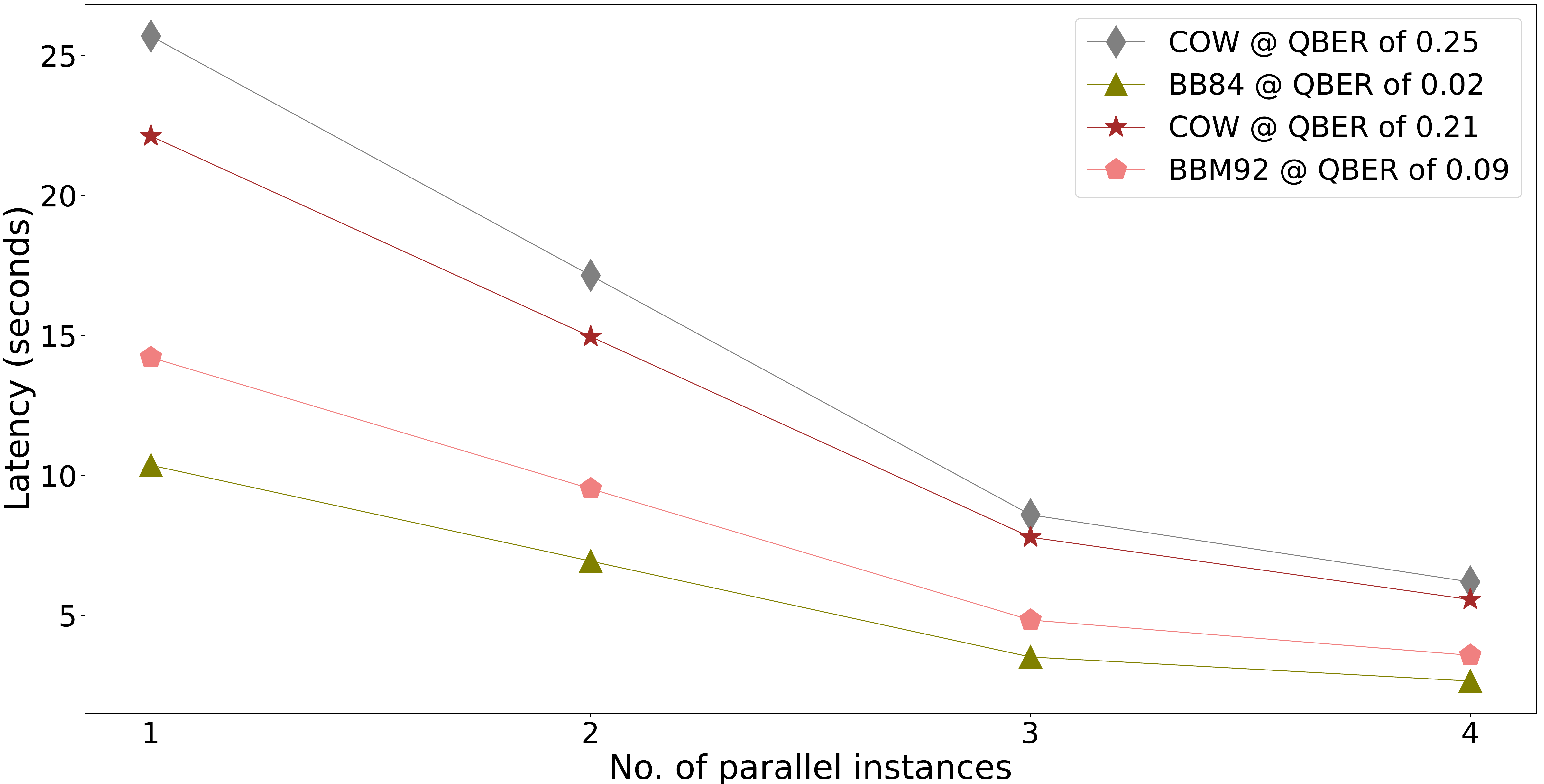}
 \includegraphics[width=0.5\textwidth]{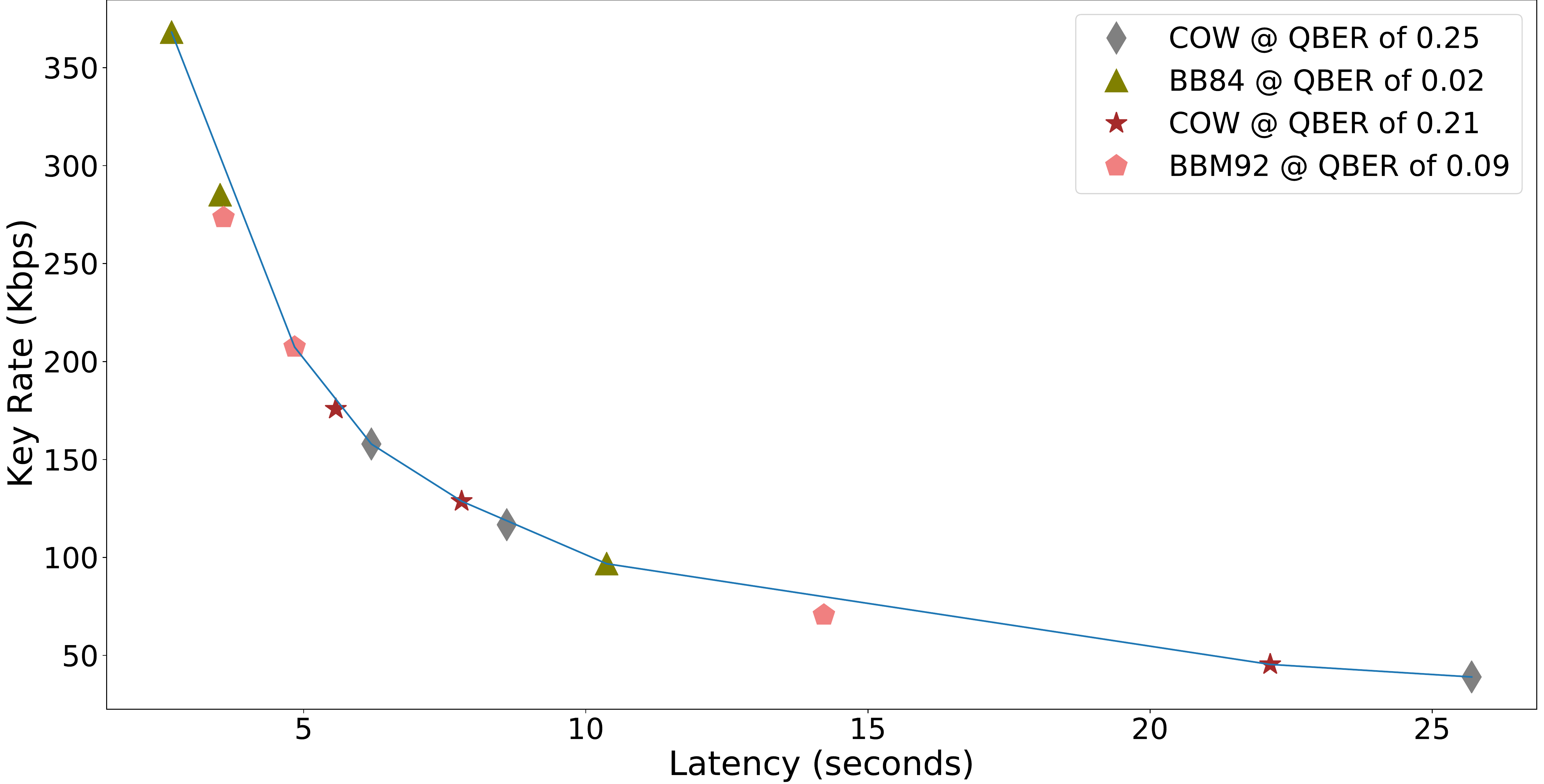}
\caption{Plot of (a) Number of parallel instances of mapper in the Hadoop framework vs execution time in seconds, and (b) Execution time, in seconds, vs Key rate in Kbps, for COW, BB84, and BBM92 QKD protocols at varying QBER (measured during the quantum experiment).}
\label{plot}
\end{figure}

\newpage
In terms of Area, from Table. \ref{uti}, it is observed that there is an increase in the utilization of Logic Cells and Block RAM units, only on the receiver with the Hadoop framework, due to the LDPC decoder (which is computationally intensive). This can be reduced by adopting optimized decoder implementations. 
The experiment is run for each protocol, and $10Mb$ of raw key bits are collected and processed. 
The variation in QBER captured in the experiments run is to validate the ability of the rate-adaptive LDPC codes with threshold error correction capacity of 25\% at 90\% efficiency. 
In terms of Performance, it can be observed from the graph in Figure. \ref{plot} that with an increase in QBER the execution time or latency also increases and this inversely affects the rate at which the secret key can be extracted. But if we leverage the number of Hadoop parallel instances in the design, the latency can be reduced, thereby increasing the key rate. The graph thus asserts the relation: \[No.\ of\ parallel\ instance\ \propto\ QBER\],
and,
\[No.\ of\ parallel\ instance\ \propto\ \frac{1}{Latency} \]

\section{Conclusion}
In this paper, a design has been presented that strengthens the security and enables faster key reconciliation for QKD systems using the concept of FPGA-based Hadoop-map reduce architecture. A new reconfigurable architecture to optimize the resources by using reusable blocks has been proposed.  The results of our experiments show that the hardware design has a balance between resource utilization and throughput. Therefore, implementing the Hadoop-based QKD post-processing functionality  directly in the hardware is a preferred technique to meet the computation and critical security needs of commercial QKD systems. An additional advantage of FPGA based key distillation process is to improve the performance and compatibility with different types of QKD experimental setups.  Typically, QKD key distillation engine uses a combination of individual IP core blocks to build complete post-processing protocols, including a data encryption module. This makes it possible to build a secure multi-QKD protocol supporting quantum key distillation hardware with volatile FPGA systems.   
\newpage
\section{Acknowledgement}
This research was carried out with support from the grant provided by the Department of Science and Technology (DST), within the Ministry of Science and Technology, Government of India, through the “Quantum enabled Information Science and Technology (QueST)” program. The authors wish to acknowledge and thank Dr. Jothi Ramalingam and Mrs. Sarika Menon for their invaluable assistance and insightful discussions. 
\section{Author Contributions}
The theoretical aspect and ideation of the work were done by NV and FS.  Processing of quantum keys, design of the Hadoop framework, and hardware implementation of all the key distillation algorithms were done by FS, SG, and HP. The quantum experiment of all the listed QKD protocols was carried out by PC and RP. The side-channel-resistant authentication scheme was implemented by DB. The paper was written by FS and NV, and the other authors proofread it. All authors discussed the results and commented on the manuscript.


\end{document}